\documentclass[twocolumn, superscriptaddress, prb, showpacs]{revtex4}
\newif\ifGALLEYversion\GALLEYversionfalse

\usepackage{ifthen}
\usepackage{graphicx}
\usepackage{amsmath}
\usepackage{bm}
\usepackage{color}
\usepackage{ulem}

\newcommand{\rr}{{\bf r}}
\newcommand{\rrm}{{\bf r}'}

\newcommand{\kF}{k_{\text{F}}}

\newcommand{\eq}[1]{Eq.~(\ref{#1})}
\newcommand{\eqs}[2]{Eqs.~(\ref{#1}) \& (\ref{#2})}
\newcommand{\ACF}{{\text{ACF}}}
\newcommand{\RPA}{{\text{RPA}}}
\newcommand{\LDA}{{\text{LDA}}}
\newcommand{\tin}{{\text{in}}}
\newcommand{\xc}{{{xc}}}
\newcommand{\self}{\text{self}}
\newcommand{\grad}{{\mbox{\scriptsize grad}}}

\newcommand{\Tr}{\mathrm{Tr}}

\newcommand{\bq}{\mathbf{q}}

\newcommand{\bn}{\begin{equation}}
\newcommand{\ee}{\end{equation}}
\newcommand{\bga}{\begin{eqnarray}}
\newcommand{\eda}{\end{eqnarray}}

\newcommand{\diff}{\mathrm{d}}

\newcommand{\im}{\mathrm{i}}

\newcommand{\Exc}{E_{xc}[n]}

\newcommand{\nl}{\text{nl}} 
\newcommand{\vdWDF}{\text{vdW-DF}}

\newcommand{\bqm}{\mathbf{q'}}
\newcommand{\msl}{\text{sl}}

\begin{document}

\title{Interpretation of van der Waals density functionals}
\author{Per Hyldgaard} 
\affiliation{Department of Microtechnology and Nanoscience, Chalmers
University of Technology, SE-412 96 G\"oteborg, Sweden.}
\affiliation{Materials Science and Applied Mathematics, Malm{\"o} University, 
Malm{\"o} SE-205 06, Sweden.}
\author{Kristian Berland} 
\affiliation{Department of Microtechnology and Nanoscience, Chalmers
University of Technology, SE-412 96 G\"oteborg, Sweden.}
\author{Elsebeth Schr\"oder} 
\affiliation{Department of Microtechnology and Nanoscience, Chalmers
University of Technology, SE-412 96 G\"oteborg, Sweden.}
\date{\today}

\begin{abstract}
  The nonlocal
  correlation energy in the van der Waals density functional
  (vdW-DF) method [Phys.\ Rev.\ Lett.~{\bf 92}, 246401 (2004);
  Phys.\ Rev.\ B \textbf{76}, 125112 (2007); Phys.\ Rev.\ B {\bf 89},
  035412 (2014)] can be interpreted in terms of a coupling of
  zero-point energies of characteristic modes of semilocal
  exchange-correlation (xc) holes.
  These xc holes reflect the internal functional in the framework of
  the vdW-DF method [Phys.\ Rev.\ B {\bf 82}, 081101(2010)].
  We explore the internal xc hole components, showing that they share 
  properties with those of the generalized-gradient approximation.
  We use these results to illustrate the nonlocality in the vdW-DF
  description and analyze the vdW-DF formulation of nonlocal 
  correlation.
\end{abstract}

\pacs{31.15.E-, 71.15.Mb}
\maketitle

\section{Introduction}
In a seminal paper\cite{ra} Rapcewicz and Ashcroft (RA) highlighted the
connections between nonlocal correlations, the exchange-correlation (xc) 
hole concept\cite{hajo74,lape75,gulu76,lape77,gujolu79,wang91p13298} of 
density functional theory (DFT), and van der Waals (vdW) forces in the 
inhomogeneous electron gas. RA introduced a simple physical picture of 
vdW binding: electrons and their associated xc holes form neutral pairs 
in a system resembling condensed, vdW-bounded,
atomic matter and experience mutual attraction of a 
dispersive nature.\cite{eilo30,lo30,jerry65} In the RA view, 
it is the local plasmon that characterizes the interaction 
components, i.e., electron-xc-hole pairs. The RA 
picture is supported by a previous study of the nonlinear response in 
the electron gas,\cite{ma} predicting strong vdW binding from 
quantum-fluctuation contributions in the interaction diagram that also underpins 
an analysis of gradient-corrected correlation.\cite{rasolt,lape80,RasoltDisc1,hula86,RasoltDisc2} 
In fact, the long-range interaction component is interpreted\cite{lavo87}
as reflecting the small-momentum fluctuation components that are 
extracted to reach a generalized gradient approximation (GGA) in the
early formulations.\cite{lape75,lape77,lape80,lameprl1981,lape82,lameprb1983,lavo87,lavo90} 
Together Refs.\ [\onlinecite{ra,ma,lavo87}] suggest that one
can recover vdW forces in nonlocal functional theories that also
incorporate the tremendous progress that GGA 
represents.\cite{KieronPerspective,BeckePerspective}

The Rutgers-Chalmers vdW-DF method\cite{anlalu96,hurylula99,huanlula96,ryluladi00,rydbergthesis,rydberg03p126402,dionthesis,dion,dionerratum,langreth05p599,thonhauser,cooper10p161104,lee10p081101,behy13,berland14p035412,bearcoleluscthhy14,ROPPreview} 
allows efficient computations of the xc energy 
in an approximation that seamlessly incorporates nonlocal 
correlation effects, including vdW forces. The vdW-DF method is gaining 
recognition for helping to extend the success of nonempirical DFT to sparse 
matter.\cite{langrethjpcm2009} The vdW-DF method is free from external 
parameters and rests only on formal theory input,\cite{ROPPreview} for
the local density approximation\cite{gulu76,perdew92p13244} (LDA) and for 
gradient-corrected exchange\cite{lavo90,schwinger,elliott09p1485,lee10p081101} 
in its specification of the plasmon behavior. It also includes a GGA
exchange component.\cite{pewa86,pebuer96,zhya98,mulela09,cooper10p161104,vdwsolids,berland14p035412,hamada14} 
The choice of vdW-DF exchange can be guided by conservation of the full 
(nonlocal) xc hole\cite{berland14p035412} and with such consistent-exchange vdW-DF it is possible 
to investigate bulk-structure and adsorption problems where interactions are in subtle 
competition.\cite{berland14p035412,bearcoleluscthhy14}
The transferability of vdW-DF has also been probed via comparison with quantum Monte Carlo (QMC) 
results for hydrogen phases\cite{qmctest} and for water.\cite{qmctestwater}
The vdW-DF method was first tested in 
non-selfconsistent forms\cite{ryluladi00,rydberg03p126402,dion,langreth05p599,beloschy13} 
using GGA calculations of the electron density as input for a post-processing
evaluation of the nonlocal correlations. With a formal derivation of 
forces arising from the nonlocal correlation term\cite{thonhauser} and 
with the introduction of efficient algorithms for computing the vdW-DF 
energy and forces,\cite{linearscaling,roso09} the vdW-DF method today benefits 
from experience in widespread sparse- and general-matter
applications.\cite{langrethjpcm2009,bearcoleluscthhy14,ROPPreview,cooper08p1304,berland10p134705,londero11p054116,vdwsolids,berland11p1800,ni12p424203,ihm12p424205,lazic12p424215,vdWReady,Mperspectives,poloni12p4957,poloni14p861,bjorkman14}

The nonempirical vdW-DF-method is built around a semilocal internal 
(or inner) 
functional,\cite{lee10p081101,berland14p035412,bearcoleluscthhy14,ROPPreview} 
with xc hole $n_{xc}^{\tin}$,
and an evaluation of a nonlocal correlation energy $E_c^{\nl}$.
The internal functional keeps local exchange and correlation together but
limits gradient corrections to exchange.  The internal functional 
was introduced as a concept in Ref.\ [\onlinecite{lee10p081101}] but it 
underpins all formulations of the general vdW-DF 
versions\cite{rydbergthesis,dionthesis,dion,ROPPreview} 
since it serves to model the local variation in the plasmon-pole 
response from which $E_c^{\nl}$ is formulated. There have to
date only been brief discussions of this internal 
functional.\cite{rydbergthesis,lee10p081101,bearcoleluscthhy14} 

The goal of this paper is to present the vdW-DF construction formally both in terms of the internal xc hole 
and physical pictures.  Standard vdW-DF 
presentations\cite{dion,langreth05p599,thonhauser,lee10p081101,bearcoleluscthhy14}
start more directly with a plasmon-pole representation of the 
response. However, there are benefits of tracing the plasmon view back 
to a discussion of the associated internal functional xc holes. This 
makes it possible to discuss the close connection that exists between
vdW-DF and the GGA descriptions. Moreover, the emphasis on the 
internal-functional response allows us to interpret the vdW-DF 
nonlocal correlation energy in terms of the RA physics picture of vdW 
forces.\cite{ra} In turn these results allow us to illustrate the mechanisms
by which vdW-DF retains a collectivity and nonlocality in its description
of the screened response and materials interactions.

The paper is organized as follows. In section \ref{sec:vdWDFdesignstrategy}, 
we present an xc-hole based formulation of the vdW-DF framework. 
It is meant to give the reader an overview
of the vdW-DF method in a self-contained and
alternative derivation cast in the concepts that we
explore in this paper.
In section III we plot and discuss the internal functional
xc-hole components of vdW-DF. Section IV contains a demonstration of 
the link between the vdW-DF nonlocal correlation energy
and the RA physics picture. Finally, section V 
summarizes the paper, while an appendix details that the vdW-DF 
rests on a correct longitudinal projection in its description of the 
electrodynamics coupling.
\section{The vdW-DF framework}
\label{sec:vdWDFdesignstrategy}

\begin{figure}
\includegraphics[width=\columnwidth]{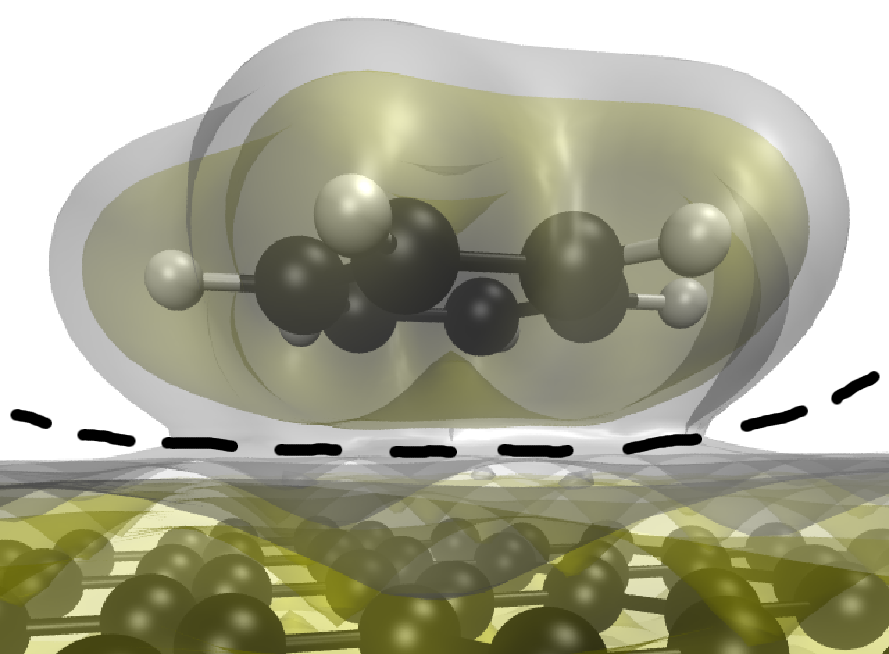}
\caption{Schematics of a typical problem where vdW-DF is called 
upon to describe the material binding: a system with multiple molecular-type 
regions that couple electrodynamically across
internal `voids' with sparse electron distribution,
Ref.\ [\protect\onlinecite{langrethjpcm2009}].  Each molecular-type
fragment can be described by a GGA-type description and such
descriptions form the starting point for the vdW-DF evaluation
of nonlocal correlations.  The `voids' are not necessarily free
of electrons but are regions in which there is only a tail
(or overlapping tails) of the molecular-type electron
distributions. The schematics is adapted from Fig.\ 7 of
Ref.\ [\protect\onlinecite{behy13}] and shows the atomic
configuration and contours of the electron-density
distribution of the molecular building blocks for a
benzene molecule and a graphene sheet at the vdW-DF
binding separation.  We note that a natural delineation
surface of minimum electron distribution (here illustrated by
a dashed curve) runs through inter-fragment positions
with a saddle-point or trough-like behavior in the electron concentration.
}
\label{fig:delineation}
\end{figure}

To discuss the nature of vdW-DF, we start out by noting that constraint-based GGA has been 
enormously successful at describing chemically connected systems, both those that have 
zero dimensions (such as atoms, molecules, and nanoparticles) and those with one or more 
macroscopic dimensions (such as wires, surfaces, sheets, and solids).\cite{KieronPerspective,BeckePerspective}
We denote such systems `molecular-type' even if they can be infinitely extended. 

Our discussion benefits from considering two or 
more such molecular-type regions in close vicinity. 
Fig.~\ref{fig:delineation} shows an example 
with benzene adsorbed on graphene at a separation of 3 {\AA}.\cite{behy13}
The vdW-DF method seeks to extend the GGA success (for an individual 
fragment) by adding an account of the nonlocal correlations that arise 
among several such fragments as well as inside the 
fragments.\cite{langrethjpcm2009,bearcoleluscthhy14,vdwsolids}
Only the coupling mediated by the longitudinal component of the 
electrodynamical interaction described in the Coulomb gauge, 
$V\equiv |\rr_1-\rr_2|^{-1}$ is considered.  The Coulomb Green 
function is $G=-4\pi V$.  

The starting point is the exact adiabatic 
connection formula\cite{lape75,gulu76,lape77} (ACF):
\begin{eqnarray}
E_{xc} +E_{\self}
& = & - \,\int_0^1 d\lambda \, \int_0^{\infty} \frac{d\omega}{2\pi} \,
        \text{Tr}\{ \Im \chi_\lambda(\omega) V \} \, ,
        \nonumber\\
 & = & - \int_0^1 d\lambda \, \int_0^{\infty} \frac{du}{2\pi} \,
        \text{Tr}\{ \chi_\lambda(\im u) V \} \, ,
        \label{eq:ACFexpress}
\end{eqnarray}
where
$\lambda$ is the electron-electron coupling constant and 
 `Tr' denotes a full trace over the variation\cite{notationfootnote}  
in the reducible density-density correlation function 
$\chi_{\lambda}(iu)$ at imaginary frequency $u$. 
The infinite self energy is given by 
$E_{\self} =(1/2) \text{Tr}\{ V \hat{n}\}$ where
$\hat{n}(\rr)$ is the density operator.  The reducible
density-density correlation function relates an external 
potential change $\delta \Phi_{\rm ext}$, at a characteristic
frequency $\omega$, to resulting density changes, $\delta n = 
\chi_\lambda(\omega) \delta \Phi_{\rm ext}$. The electron 
dynamics causes any such external potential to be screened;
the system can also be described by a corresponding screened 
potential  $\delta \Phi_{\rm scr}$. The irreducible density-density 
correlation function $\tilde{\chi}_0$ relates this screened 
potential to the same density change, $\delta n = 
\tilde{\chi}_\lambda(\omega) \delta \Phi_{\rm scr}$.
A Dyson equation relates $\tilde{\chi}_\lambda$ and
$\chi_\lambda$, Ref.\ [\onlinecite{lape77}].

The vdW-DF framework expresses the xc energy
\begin{equation}
E_{xc} + E_{\self}  =   \int_{0}^{\infty} \, \frac{du}{2\pi}
\hbox{Tr} \{ \ln ( \kappa_{\ACF}(\im u)) \} \, ,
\label{eq:FirstExcRecast} 
\end{equation}
in terms of an approximation for an effective longitudinal 
dielectric function $\kappa_{\ACF}(\im u)$. We make three observations.
First, the expression (\ref{eq:FirstExcRecast}) is formally equivalent to the
ACF,\cite{lape75,gulu76,lape77} as given in \eq{eq:ACFexpress}, since the 
coupling-constant integration is captured in the definition of $\kappa_{\rm ACF}$. 
Second, the explicit relation is given in terms of an effective  external-potential 
(density) response function $\chi_{\ACF}(\omega)$ via
\begin{eqnarray} 
        - \chi_{\ACF}(\omega) V  \equiv 
1- \exp\left[-\int_0^1 \, d\lambda \, S_{\lambda}(\omega) \right] \, ,
\label{eq:xacfdef}
\end{eqnarray}
where $S_\lambda(\omega) \equiv - \chi_{\lambda}(\omega)V$ denotes the
fluctuation or plasmon propagator\cite{pinesnozieres,lu67} at coupling 
constant $\lambda$.  The corresponding longitudinal dielectric function is
\begin{equation}
        \kappa_{\ACF}(\im u) \equiv  [1+\chi_{\ACF}(\im u) V]^{-1} 
         = \exp\left[\int_0^1 \, d\lambda \, S_{\lambda}(\omega) \right]\, .
        \label{eq:trueresponserelation}
\end{equation}
Third, writing 
\begin{equation}
\kappa_{\ACF}(\im u)=\nabla \epsilon(\im u) \cdot \nabla G \, ,
\label{eq:kappaset}
\end{equation}
makes $\kappa$ a rigorously defined longitudinal projection
of an effective dielectric function $\epsilon$, as further detailed in
appendix \ref{sec:longproj}.  In vdW-DF it is assumed that a 
scalar, but nonlocal, dielectric function, $\epsilon(\im u)$ can 
be used in \eq{eq:kappaset}; there exists a demonstration that such 
a scalar $\epsilon$ can be constructed for any given xc energy functional 
(for example, the exact functional).\cite{rydbergthesis,ROPPreview}
 
To specify the plasmon-pole description and the xc hole components
we also introduce an effective screened (density) response function 
$\tilde{\chi}_{\ACF}$ given by
\begin{eqnarray}
        \tilde{\chi}_{\ACF}(\im u)V & \equiv & \kappa_{\ACF}(\im u) \, 
\chi_{\ACF}(\im u)V 
        \label{eq:DysonACF} \\
 & = & 
\chi_{\ACF}(\im u)V \,\kappa_{\ACF}(\im u) \nonumber\\
& = & 1-\exp[\int_0^1 d\lambda \, S_\lambda]
\, .
        \label{eq:DysonACFeval}
\end{eqnarray}
This definition complies with the Lindhard-type formulation\cite{lindhard}
 \begin{equation}
        \kappa_{\ACF}(\im u) \equiv  1 - \tilde{\chi}_{\ACF} V \, .
        \label{eq:lindhardrelation}
\end{equation}
The screened response \eq{eq:DysonACF} is specified via a longitudinal projection
\begin{equation}
        \tilde{\chi}_{\ACF}(\im u)  = \nabla \alpha(\im u) \cdot \nabla 
        \label{eq:tilchiacfalpharelation} 
\end{equation}
of the local-field dielectric response $\alpha(\im u)$. From \eq{eq:lindhardrelation}
it is clear that a scalar approximation for $\alpha$ also specifies the 
vdW-DF dielectric functional, $\epsilon \equiv 1 + 4\pi \alpha$, that 
enters in \eq{eq:kappaset} and determines the xc energy \eq{eq:FirstExcRecast}
in the vdW-DF framework.

The many-body response nature of any xc functional is naturally 
expressed in the ACF evaluation of the xc hole
\begin{equation}
        n_{xc}(\mathbf{r};\mathbf{r}')= -\frac{2}{n(\mathbf{r})}\,
        \left[\int_0^{\infty} \frac{\diff u}{2\pi} \int_0^1 \diff \lambda \, 
        \chi_\lambda(\im u; \mathbf{r},\mathbf{r}')\right] - \delta(\mathbf{r}-\mathbf{r}'),
\label{eq:holeresponserelation}
\end{equation}
the electron-deficiency (at $\mathbf{r}'$) produced around an electron at point $\mathbf{r}$.

For a description of the density functional it is sufficient to
work with the spherically averaged xc hole
\begin{equation}
\bar{n}_{xc}(\mathbf{r}; r'') = \frac{1}{4\pi (r'')^2} 
\int_{|\mathbf{r}'-\mathbf{r}|=r''} \diff \mathbf{r}'\, n_{xc}(\mathbf{r};\mathbf{r}').
\label{eq:sphereaverage}
\end{equation}
The local xc energy per particle $\varepsilon_{xc}(\mathbf{r})$ is directly
related to this xc hole
\begin{eqnarray}
        E_{xc} & \equiv & \int \diff \mathbf{r} \, n(\mathbf{r}) \, \varepsilon_{xc}(\mathbf{r}) \, ,
        \label{eq:varepsidef}\\
        \varepsilon_{xc}(\mathbf{r}) & \equiv &
        \frac{1}{2} \int \diff \mathbf{r}' \,
        \frac{n_{xc}(\mathbf{r};\mathbf{r}')}{|\mathbf{r}-\mathbf{r}'|} \nonumber\\ 
        & = &  \frac{1}{2}\int_0^{\infty} r''\, \diff r''\, \bar{n}_{xc}(\mathbf{r}; r'') \, .
        \label{eq:nxcform}
\end{eqnarray}
The exact relation
\begin{eqnarray}
        E_{xc} & = & \frac{1}{2} \int \diff \mathbf{r}\, n(\mathbf{r}) \int_0^{\infty} r''\, \diff r''\, \bar{n}_{xc}(\mathbf{r}; r'')\nonumber\\
        & = & \int_0^{\infty} \frac{\diff u}{2\pi} \, \hbox{Tr} \{ \ln(\kappa_{\ACF}(\im u)) \} - E_{\self}
        \label{eq:varepsievalA}
\end{eqnarray}
links $\langle \mathbf{r} | \ln(\kappa_{\ACF}) | \mathbf{r} \rangle$ to $\varepsilon_{xc}(\mathbf{r})$,
and hence to an integral over the xc hole.

\subsection{The vdW-DF logic}

A central idea in the vdW-DF framework is to exploit that the formally exact 
formulation \eq{eq:kappaset} has already made one instance of the 
electrodynamics coupling $V\propto G$ explicit.  One obtains
truly nonlocal effects in the xc functional even when using a semilocal 
GGA-type functional to specify the details of the nonlocal 
form of $\epsilon$.  Accordingly, in the vdW-DF method we split the total 
xc energy functional and associated xc holes into semilocal and 
nonlocal contributions,\cite{rydbergthesis}
\begin{eqnarray}
\Exc & = & E_{xc}^{\msl}[n] + \Delta E_{xc}^{\nl}[n] \, ,
\label{eq:ExcGenSplit} \\
n_{xc}(\mathbf{r};\mathbf{r}'-\mathbf{r}) & = & n_{xc}^{\msl}(\mathbf{r};\mathbf{r}'-\mathbf{r}) 
+ \Delta n_{xc}^{\nl}(\mathbf{r};\mathbf{r}'-\mathbf{r}) \, .
\label{eq:nxcsplit} 
\end{eqnarray}
The first term $E_{xc}^{\msl}[n]$ of \eq{eq:ExcGenSplit} is also
called the outer semilocal functional and it is given by LDA correlation and 
a GGA description of gradient-corrected exchange. It would in principle 
provide an approximate description of a typical GGA 
problem (i.e., an individual of the molecular-type fragment shown
in Fig.\ \ref{fig:delineation}) because gradient-corrections to 
exchange are typically more important than gradient-corrected 
correlation.\cite{pecosabu06,PBEsolhole} We shall, for ease of 
discussion, also refer to such a description as being of a GGA 
type. At the same time we note that $E_{xc}^{\msl}$
should not be evaluated in isolation.

The second term $\Delta E_{xc}^{\nl}$ is viewed as a 
perturbation,\cite{rydbergthesis,ROPPreview} 
capturing nonlocal correlation energy from the coupling of plasmon poles 
that characterize $E_{xc}^{\msl}$. The formulation of $\Delta E_{xc}^{\nl}
\approx E_c^{\nl}$ is, however, in practice based on the use of an
internal semilocal functional $E_{xc}^{\tin}[n]$ that is similar 
to $E_{xc}^{\msl}[n]$, but with an energy per particle $\varepsilon_{xc}$
that decreases more rapidly at large values of the scaled density
$s=|\nabla n|/(6\pi^2 n)^{1/3}/n$. This choice 
is made to avoid spurious contributions emerging from 
low-density regions.\cite{ra,anlalu96,dionthesis,berland14p035412,bearcoleluscthhy14,ROPPreview}
The construction via a GGA-type $E_{xc}^{\tin}$ (that also just contains
LDA correlation plus GGA gradient-corrected exchange) allows vdW-DF to 
rest exclusively on formal diagrammatic input\cite{perdew92p13244,rasolt,lavo87,ra}
while avoiding\cite{thonhauser} to explicitly formulate a gradient-corrected 
correlation term $\delta E_c^{\rm grad}$ which is a necessary but also 
complex step 
in the GGA formulations.\cite{lape80,lameprl1981,lavo87,pebuwa96,pebuer96}

The vdW-DF internal functional $E_{xc}^{\tin}$ is given by a GGA-type
internal-functional xc hole $n_{xc}^{\tin}\approx n_{xc}^{\msl}$ and it is 
used to introduce an approximate scalar dielectric function $\epsilon$ via
\begin{equation}
E_{xc}^{\tin} + E_{\self} 
= \int_0^{\infty} \, \frac {\diff u}{2\pi} \, \hbox{Tr} 
	\{ \ln(\epsilon(\im u))|_{\grad}\} \, .
\label{eq:Exc0form}
\end{equation}
This form is motivated by the observation that the 
longitudinal projection in \eq{eq:kappaset} becomes redundant in the 
homogeneous electron gas (HEG) limit. As indicated by the subscript `grad',  
the $\epsilon$ definition via \eq{eq:Exc0form} rests on an 
expectation\cite{rydbergthesis,ryluladi00,langreth05p599} 
that this simplification holds approximately true for a weakly 
perturbed electron gas.  Effectively, we write 
\begin{eqnarray}
	\epsilon(\im u) & = & \exp[S_{xc}(\im u)] \, ,
\label{eq:ACFequivalentSlambda} \\
E_{xc}^{\tin} + E_{\self} 
& = & \int_0^{\infty} \frac {\diff u}{2\pi} \hbox{Tr} 
        \{ S_{xc}(\im u) \} \, .
\label{eq:Exc0iguide}
\end{eqnarray}

The vdW-DF dielectric function (\ref{eq:ACFequivalentSlambda})
is used, via \eq{eq:tilchiacfalpharelation}, to also determine 
an approximation for the full dielectric function $\kappa_{\ACF}$
and hence extend the account to also include nonlocal correlations. 
\eq{eq:Exc0iguide} is a GGA-guided ansatz for $S_{xc}$ 
(and hence $\epsilon$) that describes the effective 
full coupling-constant integration and screening effects within vdW-DF. 
We note that $S_{xc}=\ln(\epsilon)$ coincides to linear order with
the related approximation $S(\omega) \equiv 1-\epsilon^{-1}(\omega)$
that was used in the vdW-DF method presentation by 
Dion~{\it et al.}\cite{dion}  Refs.\ 
[\onlinecite{rydbergthesis,dion,ROPPreview}] suggest explicit forms
for $S_{xc}$ (and $S$), given in terms of a model plasmon 
dispersion $\omega_q(\rr)$ at two coordinate points.  

Using $S(iu)= 1-\exp[-S_{xc}] \approx S_{xc}(iu)$, we 
interpret the poles $S_{xc}$ as an approximative specification
of the collective modes $\omega_\eta$ of the system described by the
internal functional \eq{eq:Exc0form}, i.e., the zeros of 
$\det|\epsilon(iu)|$. In the HEG limit the plasmon-pole
dispersion $\omega_q$ (entering $S_{xc}$) is the same everywhere
and renders a direct specification of $\omega_{\eta}$; in the
presence of gradients, the spatial and momentum variation in 
the $S_{xc}$ plasmon poles, $\omega_q(\rr)$, represent
instead only an approximative specification of the set of 
internal-functional collective modes $\{\omega_{\eta}\}$.  
In any case, these plasmon modes are a direct reflection 
of the shape of the semilocal internal-functional xc hole, 
as explained in Sec.\ \ref{subsec:internalfunctional}.

The general-geometry vdW-DF versions\cite{dion,lee10p081101,berland14p035412} 
approximate the xc energy 
\begin{eqnarray}
E_{xc}^{\vdWDF}[n] & = & E_{xc}[n]  - \delta E_{xc}[n] \, ,
\label{eq:ExcSplit} \\
& = & E_{xc}^{\msl} + E^{\nl}_c\, ,
\label{eq:ExcvdWDFSplit} 
\end{eqnarray}
where the vdW-DF nonlocal correlation term
\begin{eqnarray}
E_{c}^{\nl} & \equiv & E_{xc}-E_{xc}^{\tin} = E_{xc}^{\vdWDF}-E_{xc}^{\msl} \nonumber \\
& = & \int_0^{2\pi} \, \frac{\diff u}{2\pi} \, 
\left[ \hbox{Tr}\{\ln(\kappa_{\ACF}(\im u)) - \ln(\epsilon(\im u)) \} \right] \, ,
\label{eq:nonlocfunctionaldef}
\end{eqnarray}
is evaluated by expanding both terms in the same plasmon-pole
description $S_{xc}$:
\begin{equation}
  E_{c}^{\nl}=\int_0^\infty \frac{\diff u}{4\pi} \Tr \left[S_{xc}^2 - \left( \nabla 
S_{xc} \cdot \nabla G \right)^2  \right] \, .
\label{eq:Enlc_Slcnew}
\end{equation}
This quadratic expansion for $E_c^{\nl}$ has
the same appearance whether cast in $S_{xc}$ (as
done in Refs.\ [\onlinecite{rydbergthesis,ROPPreview}]) or in terms of
$S$ (as done in Ref.\ [\onlinecite{dion,langrethjpcm2009}]) 
because these agree to lowest order.

For given choices of the internal-functional form (and hence of plasmon poles
in $S_{xc}$, Ref.\ \onlinecite{rydbergthesis,ROPPreview}) and of 
$E_{xc}^{\msl}$ the vdW-DF form generally discards a cross-over term 
\begin{equation}
\delta E_{xc} = 
E_{xc}^{\tin}
-
E_{xc}^{\msl}
\, .
\label{eq:deltaExdef}
\end{equation}
An improved alignment between $E_{xc}^{\tin}$ and $E_{xc}^{\msl}$ 
minimizes the difference between $E_{xc}^{\vdWDF}$ 
and an evaluation based on the formal ACF recast \eq{eq:FirstExcRecast}. 
Such an alignment reflects consistency\cite{berland14p035412} between the 
plasmon response of the internal functional and that which characterizes 
$E_{xc}^{\msl}$, and it is beneficial because it allows the longitudinal projection 
[in \eqs{eq:FirstExcRecast}{eq:kappaset}] to leverage an automatic conservation
of the full xc hole.\cite{berland14p035412,bearcoleluscthhy14} 

\subsection{The vdW-DF internal functional specification}
\label{subsec:internalfunctional}

The internal functional is semilocal and of a GGA type. It is specified
by LDA exchange energy per particle, $\varepsilon_{x}^{\LDA}(\rr) 
=-(3/4\pi)k_F(\rr)$, where $k_F(\rr)=(3\pi^2n(\rr))^{1/3}$ denotes
the local Fermi wavevector, and an enhancement factor, 
\begin{equation}
\varepsilon_{\xc}^0(\mathbf{r}) = 
\varepsilon_x^{\LDA}(\rr) \, f_{\xc}^{\tin}(n,s) \,.
\label{eq:innervarepsilon}
\end{equation}
The internal functional is thus fully given by the local 
value of the density $n(\rr)$ and of the scaled density gradients, 
$s(\rr)=|\nabla n(\rr)|/2n(\rr)k_F(\rr)$.  In the
vdW-DF design,\cite{dion,thonhauser} the internal
functional is exclusively given by 
the LDA-correlation term $f^{\LDA}_c(n)$ (independent 
of $s$) and an exchange gradient enhancement $f_x^{\tin}(s)$ 
(independent of $n$):
\begin{equation}
f_{\xc}^{\tin}(n,s) = f^{\LDA}_c(n)+f_x^{\tin}(s) \, .
\label{eq:enhancementsplit}
\end{equation}
In the vdW-DF1\cite{dion,dionerratum} and vdW-DF-cx\cite{behy13} versions we 
stick with the Langreth-Vosko analysis for screened exchange,\cite{lavo90} 
giving 
\begin{equation}
f_x^{\tin} = 1 - \, \left(\frac{Z_{ab}}{9}\right) \, s^2 \, ,
\label{eq:fgradexchangeform}
\end{equation}
specified by $Z_{ab}=-0.8491$. In vdW-DF2,\cite{lee10p081101} formal scaling 
analysis\cite{schwinger,elliott09p1485} for pure exchange
yields an enhancement of curvature with $Z_{ab}=-1.887$.
The form of $f_{xc}^{\LDA}$ is taken from Ref.\ [\onlinecite{perdew92p13244}].

The resulting energy-per-particle expression (\ref{eq:innervarepsilon}) 
provides a full specification of a vdW-DF internal xc hole 
$n_{xc}^{\tin}$ inside a model that assumes 
a Gaussian spherical average form\cite{rydbergthesis,ROPPreview}
\begin{equation}
\bar{n}^{\tin}_{\xc}(\rr,q) = - \exp[-\gamma (q/q_0(\rr))^2]\,.
\label{eq:gaussian}
\end{equation}
The simple form enables analytical evaluation for many of the spatial 
integrations in the resulting description of $E_c^{\nl}$.
Also, the model form \eq{eq:gaussian} ensures  
that $n_{\xc}(\rr,q)$ is itself conserved,
\begin{equation}
\bar{n}_{xc}^{\tin}(\mathbf{r},q\to 0) = -1 \, ,
\label{eq:formalHeval}
\end{equation}
for all exchange-enhancement choices in $f_{\xc}^{\tin}$ 
and for any value of $\gamma$.  Choosing $\gamma=4\pi/9$ in 
the Gaussian model \eq{eq:gaussian} provides a simple relation 
between the inverse length scale\cite{dion}
\begin{equation}
q_0[n](\rr) = \kF[n](\rr) \, f_{\xc}^{\tin}[n](\rr) 
\label{eq:q0relation}
\end{equation}
of the model hole $n_{xc}^{\tin}$ and the internal functional 
energy-per-particle variation \eq{eq:innervarepsilon}.  This variation
in $q_0$ is in turn used to formulate the vdW-DF evaluation of $E_c^{\nl}$ 
in terms of a universal kernel $\phi_{\nl}$, as detailed in
Refs.\ [\onlinecite{rydbergthesis,dion,dionerratum,thonhauser,ROPPreview}].

An important point for our discussion and interpretation, 
Sec.\ \ref{sec:zeropointcount}, is that the shape of the internal 
semilocal xc hole, given by \eqs{eq:gaussian}{eq:q0relation}, is used 
in vdW-DF to determine the local variation in the plasmon poles $\omega_q(\rr)$, 
Refs.\ [\onlinecite{rydbergthesis,dion,ROPPreview}]. The connection is
made by noting that the spherical averaged xc hole $\bar{n}_{\xc}^{\tin}(\rr,q)$ also defines a
natural wavevector decomposition\cite{lape75,lape77,lape80,lavo87,lavo90}
for the internal functional energy per particle,
\begin{eqnarray}
\varepsilon_{\xc}^{\tin}(\rr) & = & \int \frac{d\mathbf{q}}{(2\pi)^3} \, 
\varepsilon_{\xc}^{\tin}(\rr,\bq)\, , 
\label{eq:wavdecvareps}\\
\varepsilon_{\xc}^{\tin}(\rr,\bq) & \propto & 
\bar{n}_{\xc}^{\tin}(\rr,q)/q^2 \, .
\label{eq:varepsrel}
\end{eqnarray}
Evaluating the imaginary frequency integral in the formal relation
\begin{equation}
n(\rr) \varepsilon_{\xc}^{\tin}(\rr) = 
\int_0^\infty \, \frac{du}{2\pi} \,  
S_{xc}(iu,\rr,\rr) 
\label{eq:energydenspostcontour}
\end{equation}
with the plasmon-pole specification\cite{rydbergthesis,ROPPreview} for 
$S_{xc}(iu)$ yields a wavevector decomposition\cite{lape77}
\begin{equation}
\varepsilon_{\xc}^{\tin}(\rr,\bq) = \pi \left(\frac{1}{\omega_q(\rr)}-\frac{2}{q^2}\right) \,,
\end{equation}
that links $\omega_q(\rr)$ to the chosen description of the internal 
functional xc hole $n_{xc}^{\tin}(\rr,\bq)$.

\begin{figure}
\includegraphics[width=\columnwidth]{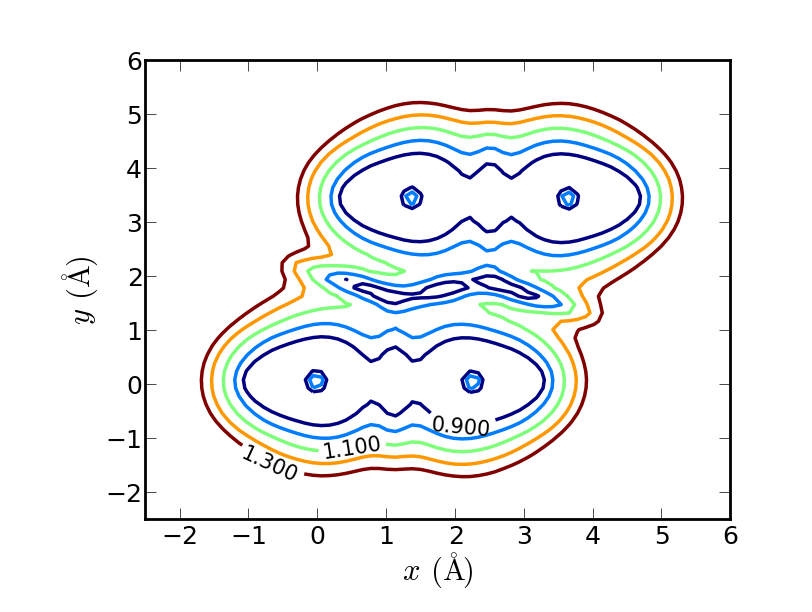}
\caption{
Contours of the scaled density gradients $s=|\nabla n|/(2k_{\rm F}n)$
in a benzene dimer at binding separation. The
saddle-point behavior, at low-density and low-to-moderate $s$ values
in the region between the molecules, is typical of binding with
a significant vdW component, Refs.\ [\onlinecite{behy13,berland14p035412}]. 
}
\label{fig:benzdimersvaluescontour}
\end{figure}
%

\section{Internal-functional exchange-correlation holes}
\label{sec:holes}

In this section, we visualize the vdW-DF internal xc hole and compare 
it to that of the numerical-GGA constructions.\cite{pewa86,pebuwa96} 
This casts light on the nature of the vdW-DF since 
these internal xc holes define the vdW-DF dielectric function 
$\epsilon$ from which vdW-DF builds an account of truly  nonlocal 
correlations. 

\begin{figure}
\includegraphics[width=\columnwidth]{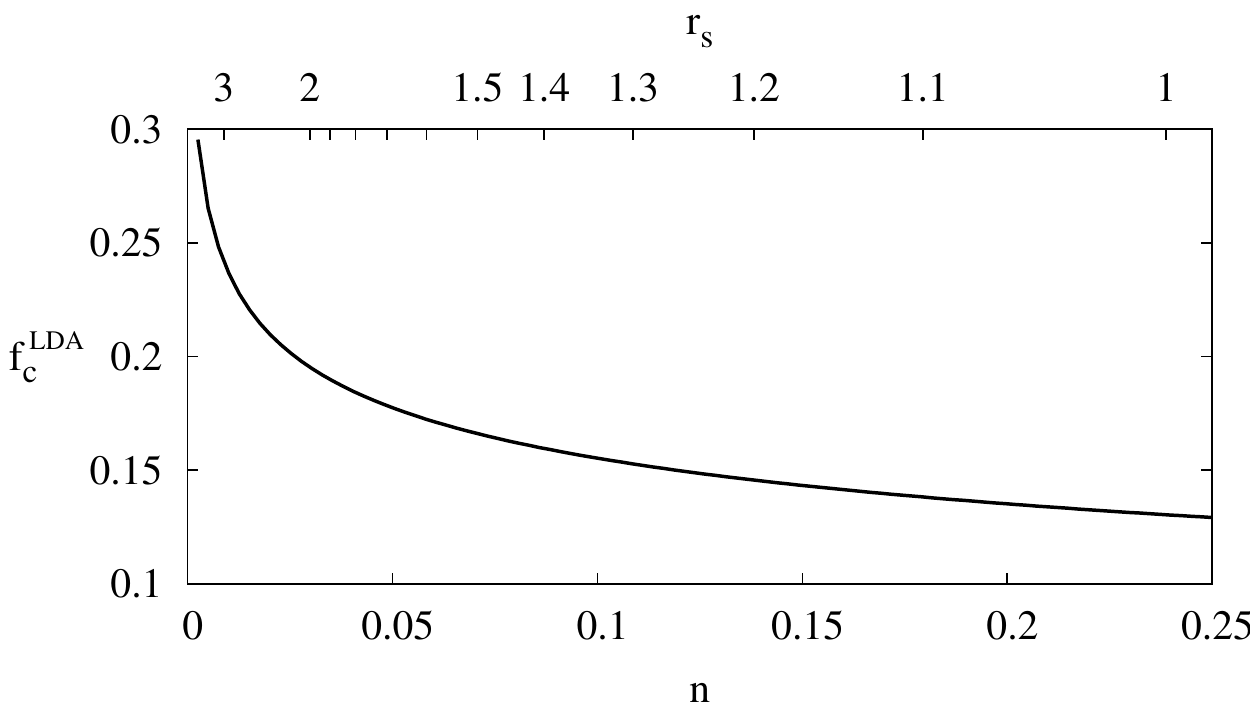}\\[1mm]
\includegraphics[width=\columnwidth]{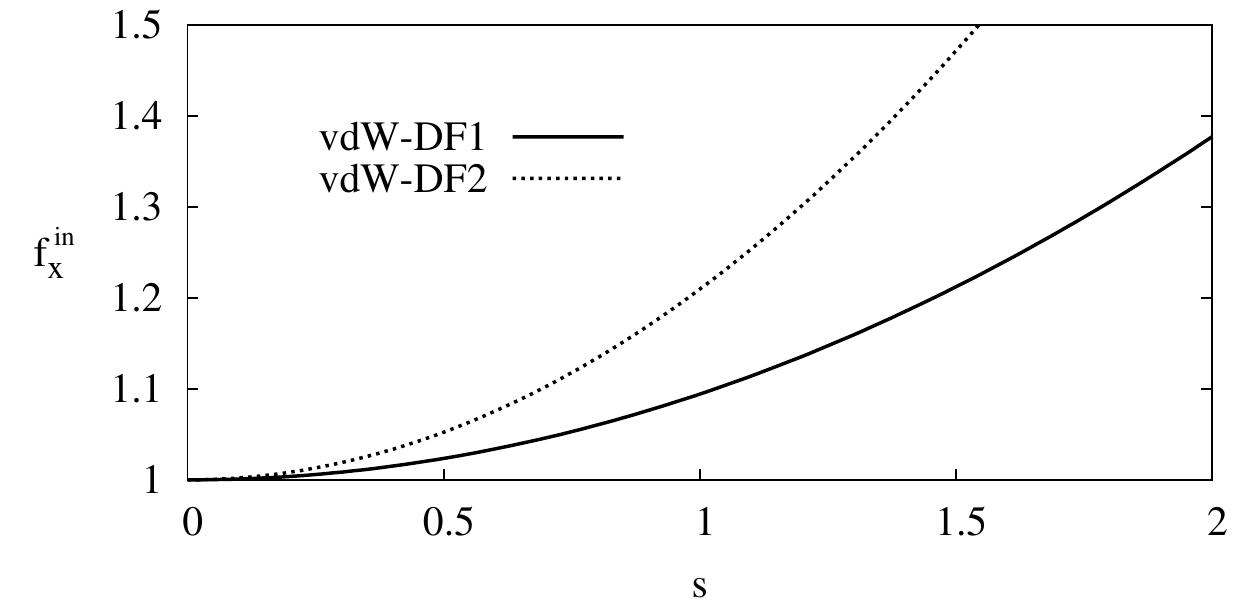}
\caption{
Correlation (top panel) and exchange (bottom panel) related components 
of the vdW-DF1 and vdW-DF2 internal functional enhancement factor 
$f_{xc}^{\tin}= f_{c}^{\rm LDA}+f_{x}^{\tin}$. The
first is a function of the density $n$ [or, equivalently,
of $r_s=(3/4\pi/n)^{1/3}$]. The second is a function of the 
scaled density gradient $s=|\nabla n|/(2k_{\rm F}n)$.
}
\label{fig:modelfxcinner}
\end{figure}

Figure\ \ref{fig:benzdimersvaluescontour} shows the
scaled-density gradient $s$ contours of a benzene dimer 
at binding separation --- a typical molecular 
binding system.\cite{berland14p035412} The plot
documents that a region with low-to-moderate  $s$ values 
exists between the molecular fragments, where the 
over-lap of two decaying densities causes a saddle-point 
or trough-like behavior.  Non-local correlation contributions arise 
from low density regions\cite{behy13} and from low-to-moderate 
values of the scaled density gradient $s$.\cite{berland14p035412} 
There is no contradiction even if $s$ typically enhances exponentially 
outside a molecular-type region, because in vdW-DF, the binding 
from nonlocal correlations arises predominantly in the 
regions between molecular 
fragments.\cite{behy13,lazic12p424215,caciuc12p424214} 

Fig.\ \ref{fig:modelfxcinner} shows the internal functional 
enhancement factor, \eq{eq:enhancementsplit} of vdW-DF1 and 
vdW-DF2. For vdW-DF1, this figure 
corresponds to density gradient $s<2$ that are often most 
relevant for $E_c^{\nl}$ binding contributions.\cite{berland14p035412}

\begin{figure}
\includegraphics[width=\columnwidth]{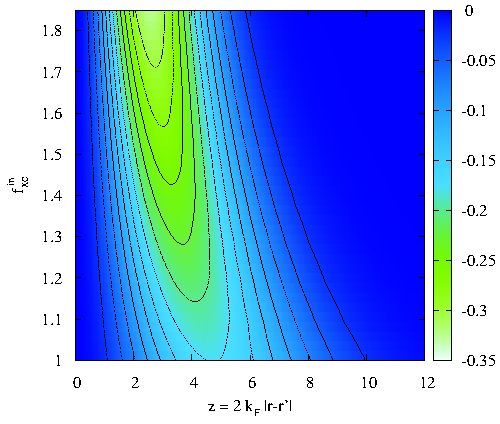}
\caption{
A contour plot of the vdW-DF internal functional 
xc hole $\bar{n}^{\rm in}_{xc}$, spatially weighted and scaled 
according to Eq.\ (\protect\ref{eq:innerfuncTothole}).
The hole is mapped as a function of scaled separation 
$z=2k_F|\mathbf{r}'-\mathbf{r}|$ and the characteristic 
internal functional enhancement factor 
$f(\rr)=f^0_{xc}(\mathbf{r})=q_0[n](\mathbf{r})/k_F[n](\mathbf{r})$
which is fixed for a given  density $n(\rr)$ and a
given scaled density gradient $s=|\nabla n|/(2k_{\rm F}n)$.
Contour spacing is 0.025. 
}
\label{fig:modelnxchole}
\end{figure}

The spherically averaged real-space xc hole 
of the internal functional of vdW-DF is extracted from 
an inverse Fourier transform of \eq{eq:gaussian}. 
This real-space internal xc hole is given by the 
following simple Gaussian form:
\begin{eqnarray}
	\bar{n}_{xc}^{\tin}(\mathbf{r};|\mathbf{r}'-\mathbf{r}|) 
        & = & - n(\rr) J(f(\rr);2k_F|\rr-\rrm|) \, ,
	\label{eq:gxc0eval} \\
         J(f,z) & = & \left(\frac{3}{2}\right)^4\, \frac{f^3}{4\pi} \,
	\exp\left(-\frac{(3 f z)^2}{64\pi}\right) \, .
	\label{eq:gxc0evalGausForm}
\end{eqnarray}
We choose to discuss the role of the internal xc hole in a form scaled 
with a distance-weighted measure:\cite{gulu76}
\begin{eqnarray}
\frac{4\pi z^2}{(2k_F)^3} \bar{n}_{xc}^{\tin}(\mathbf{r}; z) & = & 
- \left(\frac{z^2}{6\pi}\right) \, J(f(\rr); z)  \, .
\label{eq:innerfuncTothole}
\end{eqnarray}
The weighted expression (\ref{eq:innerfuncTothole}) reflects 
how the shape of the xc hole and the electrodynamics coupling $V$ 
determine the energy-per-particle variation in the 
corresponding internal semilocal functional, as given by \eq{eq:nxcform}.

Fig.\ \ref{fig:modelnxchole} shows a contour plot of 
the internal xc hole $\bar{n}_{xc}(n(\rr),s(\rr))$ as 
defined by \eq{eq:enhancementsplit} and weighted and scaled 
according to \eq{eq:innerfuncTothole}. The plot represents
the behavior of both vdW-DF1/vdW-DF-cx and vdW-DF2, since the 
vertical axis is simply the value of the internal functional
enhancement factor, $f_{xc}^{\tin}$.  The horizontal axis 
represents the scaled distance $z=2k_F|\mathbf{r}'-\mathbf{r}|$ 
from the electron position (hole center).

\begin{figure}
\includegraphics[width=\columnwidth]{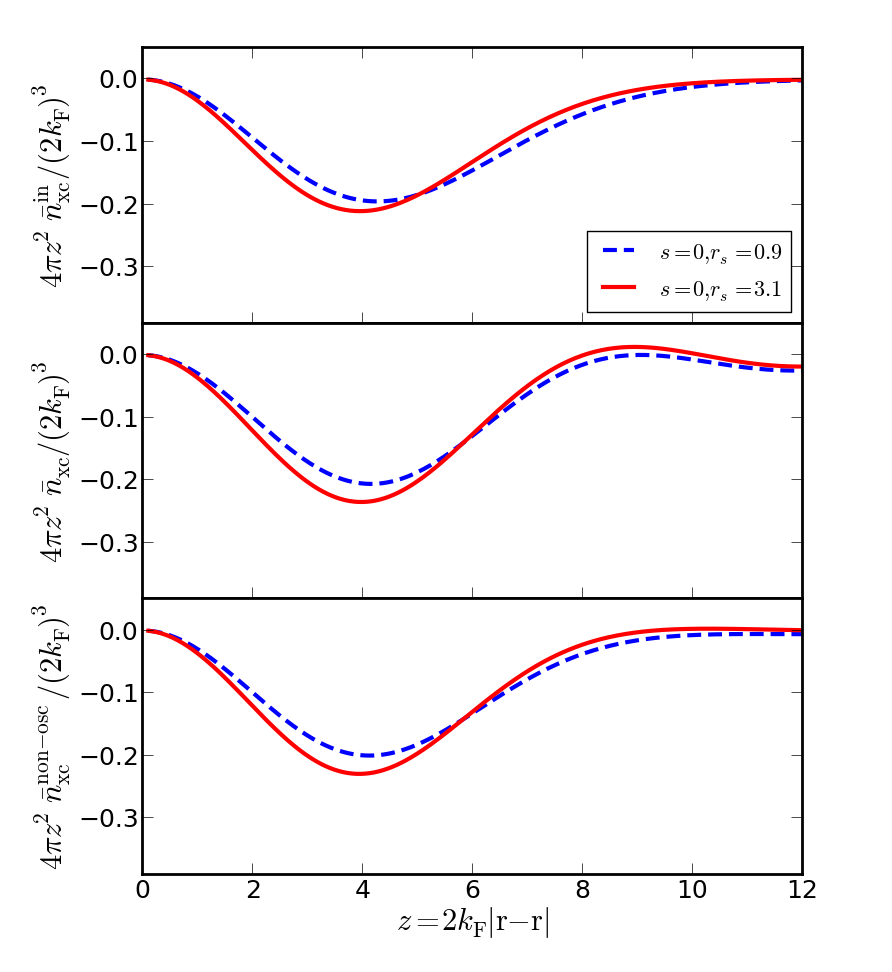}
\caption{
Comparison in the homogeneous limit between the vdW-DF representation 
of the internal xc hole (top panel) and the Perdew and 
Wang\cite{PWhole92} xc models (mid and bottom panels).  
Two electron densities, as specified by value of 
$r_s=(3/4\pi/\, n)^{1/3}$ are considered.  The mid panel 
relies on the exact exchange hole while the lower panel relies 
on a non-oscillatory approximation.  } \label{fig:LDAholecompare}
\end{figure}

Fig.\ \ref{fig:LDAholecompare} compares the internal functional xc 
hole $n_{xc}^{\tin}$ against the Perdew-Wang (PW) xc hole model for 
the homogeneous electron gas.\cite{PWhole92} This model captures 
the salient non-oscillatory features of the correlation hole and 
therefore compares well with QMC calculations.
In the bottom panel the non-oscillatory approximation for the 
exchange hole is used.\cite{PWhole92} The holes are plotted as 
functions of the scaled distances and weighted by the radial 
measure as earlier. The comparison is shown for two values of 
the Wigner-Seitz radius $r_s=(3/4\pi/\, n)^{1/3}$. The value of 
$r_s=0.9$ corresponds to the density between two neighboring C atoms of a 
benzene molecule, while the value of $r_s=3.1$ corresponds to 
the density 1.5~\AA\ out of the benzene-plane above these two 
C atoms.  

The PW model agrees well with the internal functional 
xc hole. In particular, the agreement is strikingly similar to 
the xc hole relying a non-oscillatory approximation for the 
exchange hole, with the exception of a slightly different 
trend with changing $r_s$.  In summary, vdW-DF not only
keeps a good balance between local exchange and correlation 
contributions, in line with the DFT tradition,\cite{lape77} 
but also an internal xc hole form in fair agreement with QMC.

\begin{figure}
\includegraphics[width=\columnwidth]{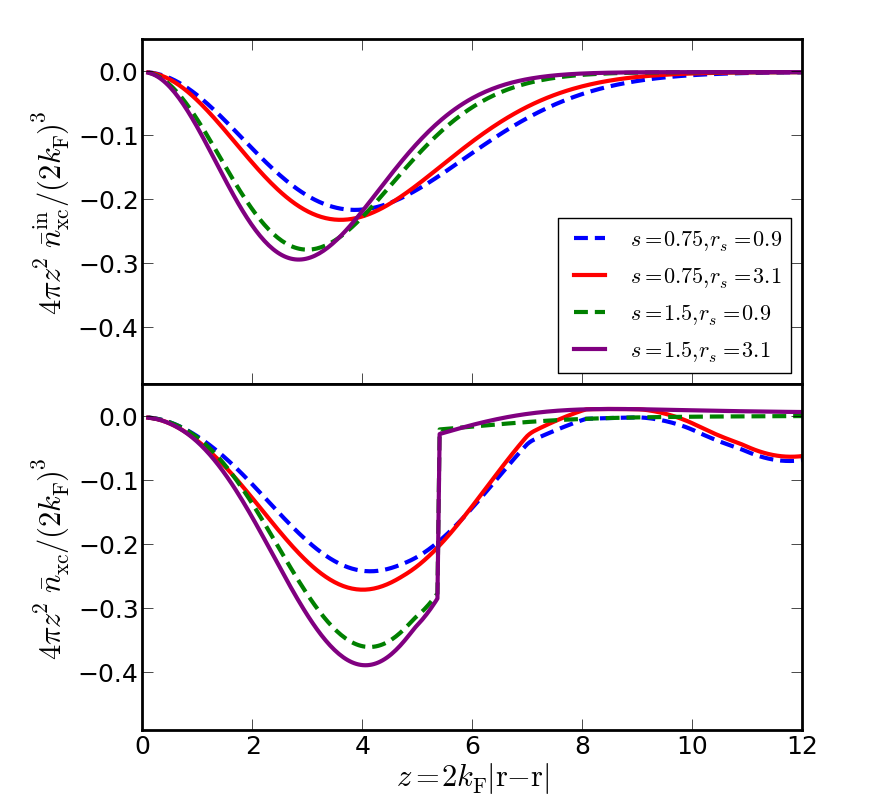}
\caption{
A comparison of the vdW-DF2 internal xc hole and a numerical 
xc hole consisting of exchange at the GGA level and correlation 
at the LDA level. The scaled gradient is chosen as $s=0.75$ and $s=1.5$ for  two different densities. The hole deepens and narrows as $s$ increases.}
\label{fig:modelnxhole}
\end{figure}

Equally interesting is the question if the vdW-DF internal-hole 
characterization also remains useful when applied to typical
systems. These have density gradients and we need to  selectively 
add the effects of gradient-corrected exchange in $n_{xc}^{\tin}$ 
(by the vdW-DF design logic).  The shape of the vdW-DF
model internal xc hole should remain reasonable at values of the 
scaled gradient $s$ that are deemed relevant for the evaluation of 
the nonlocal correlation.
 
Numerical GGA\cite{pewa86,pebuwa96,PBEsolhole} is a well defined
procedure to  impose charge conservation and a negativity 
condition on the xc hole of a gradient 
expansion\cite{rasolt,lape80,lameprl1981,lape82,RasoltDisc1,RasoltDisc2,lavo87} 
around a homogeneous electron gas.  Since this xc hole construction can be 
used to derive popular xc functionals such as PW86,  PBE, and 
PBEsol,\cite{pewa86,pebuer96,perucs08} we compare relevant 
components (including gradient corrected exchange but not 
gradient-corrected correlations) of also these to those of
the vdW-DF internal functional at relevant
nonzero values of the density gradient.

Figure\ \ref{fig:modelnxhole} compares the vdW-DF2 internal 
functional xc-hole representations at $s=0.75$ and $s=1.5$ with the 
numerical-GGA specification of Ref. [\onlinecite{pebuwa96}]. 
For the correlation, only the LDA part of the numerical xc hole is included, since only this component is used in the construction of the internal xc-hole of vdW-DF.  The vdW-DF2 version is 
chosen for comparison, because its internal functional has the same 
small-$s$ behavior as the PBE functional\cite{pebuer96} and
the numerical hole construction of Ref. [\onlinecite{pebuwa96}]
leads to a GGA xc that resembles PBE.
Conservation is built into the vdW-DF model of the internal xc 
hole (for any choice of $f_{xc}^{\tin}$,) \eq{eq:formalHeval} 
and there is no need to enforce hard cut-offs as in the numerical 
GGA construction.\cite{ROPPreview} 

Comparing the two panels of Fig.~\ref{fig:modelnxhole} with those 
of  Fig.~\ref{fig:LDAholecompare}, we see that in all cases the 
holes become deeper and shorter-ranged as $s$ increases.  However, 
whereas the two holes agree fairly well for small $s$, their shapes 
grow dissimilar as $s$ increases. The agreement with the 
numerical-GGA construction is best when (as is more relevant
for larger and flatter 
fragments close to binding separation\cite{berland14p035412}) 
we can limit the value of the scaled gradient to $s<1$. 
 
\section{Interpretation of the vdW-DF nonlocal correlation energy}
\label{sec:zeropointcount}
In this section, we show that the vdW-DF method relies on a 
nonlocal-correlation formulation that can be interpreted as an 
implementation of the RA picture of vdW forces.\cite{ra} 
As mentioned in the introduction, this picture sees nonlocal 
correlations as arising from an electrodynamical coupling of, 
originally independent,  xc holes of a traditional semilocal 
functional description; in vdW-DF represented by the GGA-type 
internal xc functional holes $n_{xc}^{\tin}$.  The vdW-DF
nonlocal correlation term  $E_c^{\nl}$ represents a
counting of coupling-induced zero-point energy shifts of 
the characteristic plasmon modes of these internal-functional 
xc holes.\cite{ra,anlalu96,ROPPreview} We provide the 
interpretation by adapting the analysis that Mahan
used to discuss the nature of vdW forces and detail
their relation to Casimir forces.\cite{capo48,jerry65}

To begin, we formulate the exact xc energy (\ref{eq:FirstExcRecast})
\bga
E_{xc}[n] + E_{\self}[n] & = & \int_{-\infty}^{\infty} \, \frac{\diff u}{4\pi}
\ln ( \Delta (y=\im u) ) \, ,
\label{eq:FirstExcRecastWdelta} \\
\Delta(y) & \equiv & \det |{\kappa}_{\ACF}(y)| \, ,
\label{eq:DeltaDetDef}
\eda
where time-reversal symmetry has been used to extend the integration 
over $u$ to $-\infty$.  The right-hand side of \eq{eq:FirstExcRecastWdelta} 
can be evaluated from a contour around the complex-frequency plane
that runs down the imaginary axis and closes around the half-plane 
of positive frequencies. This contour picks up characteristic poles
of $\det|\kappa_{\ACF}|$ with simple residues, as is evident in the
rewrite\cite{jerry65}
\begin{equation}
\frac{1}{4i\pi} \oint_c \, \frac{z\diff z}{\Delta(z)} \, 
\frac{\partial}{\partial z} \Delta (z) \, .  
\label{eq:DeltaContour}
\end{equation}
The form (\ref{eq:DeltaContour}) counts the sum of collective modes, 
given by $\Delta(y_\eta)=0$. With a general specification of 
$\kappa_{\ACF}(iu) = 1- \tilde{\chi}_{\ACF}(iu)V$  (beyond the
approximation used in the vdW-DF versions) there will also be 
corrections from poles in $\partial \Delta(z)/ \partial z$. 
This second set of poles corresponds to singularities in the 
local-field response\cite{jerry65} $\tilde{\chi}_{\ACF}$. 
Such singularities are normally associated with 
particle-hole excitations.\cite{SawadaPlas,RasoltDisc1,jerry65} 

The nonlocal correlation term \eq{eq:nonlocfunctionaldef} 
is formulated as a difference between the xc energy when defined 
in terms of $\ln\kappa_{\ACF}$ and $\ln\epsilon$. It therefore 
expresses an xc energy shift produced by an electrodynamical coupling. 
The general result \eq{eq:FirstExcRecastWdelta} allows us to discuss
the nature of this coupling-induced energy shift.  

We first consider a single molecular-type fragment, i.e., one of 
the molecular-type regions in Fig.\ \ref{fig:delineation}.  We 
use $\omega_\eta$ and $\bar{\omega}_\eta$ to denote the collective 
modes of the internal and of the full vdW-DF xc functional, given 
respectively by $\det|\epsilon(\omega_{\eta})|=0$ and 
$\det|\kappa_{\ACF}(\bar{\omega}_{\eta})|=0$. The
contour evaluation \eq{eq:DeltaContour} provides
the formal evaluation,
\begin{equation}
E_{c}^{\nl} = \frac{1}{2} \sum_{\eta} 
\left[\bar{\omega}_{\eta} - \omega_{\eta} \right] 
\, ,
\label{eq:Eval} 
\end{equation}
since, as motivated below, we can ignore contributions
from particle-hole excitations.\cite{lape77}

The vdW-DF idea of expressing all functional components
through an analysis of the response in an internal (GGA-type) 
functional is what makes \eq{eq:Eval} 
relevant for analyzing $E_c^{\nl}$.  A key observation is that the 
specification of the vdW-DF internal functional keeps local 
exchange and local correlation together to allow a 
cancellation of terms arising from particle-hole
excitations.\cite{lape77} Accordingly, Ref.\ [\onlinecite{dion}] 
uses simply a plasmon-pole representation for $S(iu)\equiv 1 -
\epsilon(iu)^{-1}$ but leaves no room for singularities directly 
in $\epsilon$.  The same observation underpins our assumption, in Sec.\ 
\ref{sec:vdWDFdesignstrategy}, that all singularities in 
the internal-functional 
specification,\cite{rydbergthesis,ROPPreview} 
$S_{xc}(iu)= \ln[\epsilon(iu)]$, should be seen as exclusively
reflecting collective (plasmon) poles of $\epsilon(iu)$.  
Moreover, while \eq{eq:trueresponserelation} ensures that
the reducible response function $\chi_{\ACF}$ has
singularities at the collective modes of $\kappa_{ACF}(iu)$,
there can be no single-particle singularities in the 
irreducible response $\tilde{\chi}_{\ACF}$. This follows 
in the vdW-DF framework (and only there) because 
$\tilde{\chi}_{\ACF}$ is set by the internal functional 
behavior, through $\epsilon=1+4\pi\alpha$ and 
\eqs{eq:ACFequivalentSlambda}{eq:tilchiacfalpharelation}. 

The formal evaluation in \eq{eq:Eval} can be used to interpret 
the vdW-DF nonlocal correlation term, \eq{eq:nonlocfunctionaldef}, 
as an implementation of the RA picture of vdW forces.\cite{ra} 
The $E_c^{\nl}$ term tracks changes in characteristic plasmon 
modes of the system as described in vdW-DF and in a 
semilocal functional defined by xc functional holes 
$n_{xc}^{\tin}$ (with a partial GGA character). Inclusion of
the electrodynamical coupling changes the dielectric functions
and hence the characteristic plasmon modes that characterize
$n_{xc}^{\tin}$. In effect, 
the energy shift in \eq{eq:Eval} tracks 
the effects of coupling the GGA-type internal 
functional holes.

It is also interesting to compare the formal framework of the vdW-DF method
and the random phase approximation (RPA).\cite{SawadaPlas,Pines,lape77,Furche08,Eshuis12,Ren12p7447}
The RPA correlation energy is\cite{lape77}
\begin{equation}
E_c^{\RPA} = \int_0^\infty \, \frac{du}{2\pi} \, 
\hbox{Tr}\{ \ln(1-\tilde{\chi}_0V)\} + \tilde{\chi}_0V\, .
\label{eq:rpacor}
\end{equation}
Our comparison will be based on the full $E_{c}^{\nl}$ representation (\ref{eq:nonlocfunctionaldef}) 
that underpins \eq{eq:Eval} and was used in an early seamless functional for layered 
structures.\cite{rydberg03p126402} We shall in Sec.~\ref{sec:discussion} return to a discussion 
of RPA and vdW-DF, keeping in mind also that the recent 
vdW-DF versions\cite{dion,lee10p081101,berland14p035412} use a second-order expansion of 
\eq{eq:nonlocfunctionaldef} in $S_{xc}$.  There are formal similarities between RPA and the 
vdW-DF method.  The vdW-DF framework, \eqs{eq:FirstExcRecast}{eq:kappaset}, builds on the exact 
ACF as does RPA; in fact, the RPA xc energy is obtained by inserting the approximation 
$\kappa_{\RPA} = 1 -\tilde{\chi}_0V$ in \eq{eq:FirstExcRecast}.  Also, the RPA correlation
energy can be exactly reformulated,\cite{SawadaPlas,Furche08}
\begin{equation}
E_c^{\RPA} = \frac{1}{2} \sum_n (\Omega_{n0} - \Omega_{n0}^{D})\, ,
\label{eq:rpacorshift}
\end{equation}
where $\Omega_{n0}^D$ ($\Omega_{n0}$) denotes a RPA excitation energy as described 
to lowest (full) order in $\lambda$. The RPA interpretation as a counting of zero-point 
energy shifts, \eq{eq:rpacorshift}, resembles the interpretation (\ref{eq:Eval}) 
that we present for the nonlocal-correlation energy in the vdW-DF method.  
There are also fundamental differences.  The RPA 
crafts $\tilde{\chi}_0$ from particle-hole excitations, typically given by Kohn-Sham 
orbitals and energies,\cite{Furche08,Eshuis12,Ren12p7447} whereas vdW-DF proceeds by 
asserting its response description through a plasmon 
model.\cite{rydberg03p126402,dion,thonhauser}  The summation in the vdW-DF
$E_c^{\nl}$ interpretation (\ref{eq:Eval}) is restricted to 
zero-point energy contributions defined by collective modes. 

Additional details about the nature of the vdW-DF nonlocal
correlation term $E_c^{\nl}$ can be obtained by considering 
the case of two molecular fragments, $A$ and $B$, separated by a 
delineation surface as illustrated in Fig.\ \ref{fig:delineation}.  In 
the following
we focus exclusively on the binding that arises in the nonlocal-correlation 
component of $E_{xc}$, noting that there will also be other
interaction components (arising through the interplay between
kinetic-energy repulsion, Coulomb terms, and in the outer semilocal 
functional $E_{xc}^0$). The analysis applies also when, as in weak 
chemisorption, there is some density overlap, as we can proceed
within a superposition-of-density scheme.\cite{harris85p1770,beloschy13} 
The analysis is not relevant for cases where there are also 
chemical bonds across the delineation surface but we defer
a discussion of limitations until we can formulate this in 
terms of criteria on the electron-response description.

To make the discussion more specific, we cast the interpretation
in terms of explicit approximations.  We let $n_{A}$ (and $n_{B}$) 
denote the density of fragment $A(B)$ when treated in isolation. 
These densities should be seen as DFT solutions as obtained
in a vdW-DF version; note that $n_A$ extends into the area 
that the delineation surfaces assign as region $B$ and vice versa.
We assume that the multicomponent density can be sufficiently 
approximated as a sum of fragment densities, $n=n_A+n_B$. This is 
a general approximation scheme\cite{harris85p1770} which is 
often accurate for systems held together by dispersive
forces in competition with other interactions.\cite{beloschy13}
From separate densities $n_{A}$ and $n_{B}$ we can define 
per-fragment screened response functions $\tilde{\chi}_{\ACF}^{*,A}$ and 
$\tilde{\chi}_{\ACF}^{*,B}$.  We note that the corresponding reducible 
response function $\chi_{\ACF}^{*,A}$ must have singularities at the vdW-DF 
collective modes $\bar{\omega}_{A}$ for fragment $A$. The same goes for 
the description of fragment $B$.
Next we introduce $\tilde{\chi}_{\ACF}^{A}$ as the 
region-projected part of this irreducible response i.e., the 
matrix formed from $\tilde{\chi}_{\ACF}^{*,A}$ by restricting 
both coordinates to reside in delineated region `$A$' 
as well as corresponding projections for the reducible 
response function $\chi_{\ACF}^{A}$ and for $\kappa_{\ACF}^{A}$. 

At this stage we can discuss the limitations on the extended
$E_c^{\nl}$ analysis presented below.
One requirement is that we approximately retain
a Dyson-like link, as in \eq{eq:DysonACF}, among the 
response descriptions even when working with the 
fragment-projected response description:
\begin{equation}
\chi_{\ACF}^{A(B)} \approx [\kappa_{\ACF}^{A(B)}]^{-1} 
\tilde{\chi}_{\ACF}^{A(B)} \, .
\end{equation}
A second, related, requirement is that the collective modes 
$\bar{\omega}_{A(B)}$ also represent the poles of 
$\chi_{\ACF}^{A(B)}$. This second requirement can be 
formulated as the condition that $\det|\kappa_{\ACF}^{*,A(B)}(iu)|$ 
(where the determinant reflects an integration over the entire 
space) has the same zeros as is found for 
$\det|\kappa_{\ACF}^{A(B)}(iu)|_{A(B)}$ (where the determinant 
range is limited to the delineated region).  The conditions can 
only hold approximately except when discussing well-separated
fragments.

Notwithstanding the requirements for using a partitioning scheme, we 
proceed to deepen our analysis of the nonlocal correlation term. 
Such a scheme has also been used, for example, to extract an asymptotically 
exact evaluation of interactions among defects on a surface supporting a 
metallic surface 
state.\cite{lako78,ei78,hype00,focus00,hyei02trio,berland09p155431,hy12p424219}
The important part of the Coulomb coupling is in this problem 
the component $V_{AB}$ of the Coulomb term that connects a point 
in the delineated region `$A$' with a point in the other region 
`$B$'. Using a simple  matrix factorization of
$\det|\kappa_{\ACF}|$ (and of $\det|\epsilon|$) we thus obtain
\begin{eqnarray}
E_{xc}^{\nl,{\rm AB}} & \approx &  \int_{-\infty}^{\infty} \, \frac{\diff u}{4\pi} \, 
\ln (\Delta^{*}(\im u)) \, ,
\label{eq:FirstFormalCoupling}  \\
\Delta^{*}(\im u)& \equiv & \det|1-\chi_{\ACF}^{A}(\im u) V_{AB} \chi_{\ACF}^{B}(\im u) V_{BA} | \, .
\end{eqnarray}

We note in passing that the result of \eq{eq:FirstFormalCoupling} is consistent 
with the traditional result for the asymptotic vdW binding\cite{anlalu96,hurylula99} 
\begin{equation}
E_{\rm vdW}(d) \equiv - \int_0^{\infty} \, \frac{\diff u}{2\pi} \, \hbox{Tr}\{
\bm{\alpha}_{\rm ext}^{A}(\im u)T_{AB} 
\bm{\alpha}_{\rm ext}^{B}(\im u)T_{BA}  \} \, ,
\label{eq:EvdWexpress}
\end{equation}
where $T_{AB}=-\nabla_{\rr_a}\nabla_{\rr_b}|\rr_a-\rr_b|$ denotes  a dipole-dipole coupling tensor between points in separate regions and where 
$\bm{\alpha}^{A(B)}_{\rm ext}$ denotes the external-field susceptibility
of fragment $A$ (or fragment $B$). The connection between 
\eq{eq:FirstFormalCoupling} 
and \eq{eq:EvdWexpress} is made by expanding the logarithm and noting that
\begin{equation}
        \chi_{\ACF}^{A(B)}(\omega) = 
\nabla \cdot \bm{\alpha}^{A(B)}_{\rm ext}(\omega) \cdot \nabla 
        \label{eq:chiacfalpharelation}
\end{equation}
specifies the vdW-DF approximation for these susceptibilities.\cite{rydbergthesis}

For a discussion of the electrodynamical coupling expressed in $E_c^{\nl}$
we provide a contour-integral evaluation of \eq{eq:FirstFormalCoupling}.
Using the contour-integration formulation (\ref{eq:DeltaContour}) for 
$\Delta^{*}$ we now have contributions
from the poles $\bar{\omega}_{\eta_{A(B)}}$ of the fragment response 
function $\chi_{\rm ACF}^{A(B)}$. We assume that these give rise
to contributions that resemble those specified by the 
exciton-susceptibility tensors in Ref.\ \onlinecite{jerry65}.  Equally important, 
the form \eq{eq:FirstFormalCoupling} has regular plasmon-pole contributions 
given by the zeros $\bar{\omega}_{(\eta_a,\eta_B)}$ of 
\begin{equation}
\Delta^{*}(\omega) = \det \left|1- \bm{\alpha}_{\rm ext}^{A}(\omega) T_{AB} 
\bm{\alpha}_{\rm ext}^{B}(\omega) T_{BA}\right| \, .
\label{eq:SecondFormalCouplingNewArgument} 
\end{equation}
Adapting the argument presented in Ref.\ \onlinecite{jerry65}, these
inter-fragment collective mode $\bar{\omega}_{(\eta_a,\eta_B)}$ correspond 
to a coupling between polarizability contributions defined in
$\bm{\alpha}_{\rm ext}^{A}(\omega)$ and $\bm{\alpha}_{\rm ext}^{B}(\omega)$
by modes $\bar{\omega}_{\eta_a}$ and $\bar{\omega}_{\eta_B}$. Overall the 
coupling contour integration leads to an approximative evaluation
\begin{equation}
E_{xc}^{\nl,AB} \approx \frac{1}{2} \sum_{{\eta_A},{\eta_B}}
\left[ \bar{\omega}_{(\eta_A,\eta_B)} - \bar{\omega}_{\eta_A} -
\bar{\omega}_{\eta_B} \right]\, ,
\end{equation}
and establishes a further link between the nonlocal-correlation term in vdW-DF 
and the RA picture, viewing vdW forces as arising as a coupling of 
(semilocal, initially independent) xc holes.\cite{ra,ma,anlalu96}
 
Finally, we note that we can extend a partition-based analysis also
to the case when there are three (or more) molecular fragments, again adopting
the analysis used for the study of electronic substrate-mediate interactions
among defects on surfaces.\cite{ei78,hyei02trio} For cases with
three molecular-type fragments, denoted $A$, $B$, and $C$, we find
\begin{widetext}
\begin{eqnarray}
        E_{\rm xc}^{\vdWDF} + E_{\rm self} & \approx &
        - \sum_{i=A,B,C} \int_0^{\infty} \, \frac{\diff u}{2\pi} \,
        \hbox{Tr} \, \{ \ln(1+\chi_{\ACF}^{i} V_{ii}) \}  
        \nonumber \\
        & & + \int_0^{\infty} \, \frac{\diff u}{2\pi} \,
        \hbox{Tr} \, \{ \ln(1 - \sum_{i<j}^{A,B,C}
        {\chi}_{\ACF}^{i} \, V_{ij} \, 
        {\chi}_{\ACF}^{j} \, V_{ji} 
        -2 {\chi}_{\ACF}^{A} \, V_{AB} \, 
        {\chi}_{\ACF}^{B} \, V_{BC} \, 
        {\chi}_{\ACF}^{C} \, V_{CA}  
        ) \} \, .
        \label{eq:partitionedACFstartrio}
\end{eqnarray}
\end{widetext}
The trio term does not naturally enter in the 
vdW-DF description when investigating a system with only two 
molecular-type regions near binding separation. The argument 
for approximating $\chi_{\ACF}^i$ as exclusively connecting two 
points inside the same fragment breaks down 
if one were to partition an individual molecular-type region.

\section{Discussion}
\label{sec:discussion}

The influence  of screening and nonadditivity effects on the vdW forces are 
explored in a significant body of literature, for example in
Refs.\ [\onlinecite{ra,Axilrod,BarashI,BarashII,SerneliusI,SerneliusII,Dobson06,Tkatchenko08,Lebegue10,Tkatchenko12,Ruzsinszky12,PT,DoGoDisp12,vdWscale,DobsonMB14,DobsonPerspective}].
In this section, we discuss to what extent the vdW-DF 
method\cite{dion,lee10p081101,berland14p035412} can capture such effects. 
In particular, we first compare vdW-DF to the RPA for the correlation energy and 
then discuss vdW-DF in light of a recently suggested classification scheme of 
dispersion interaction effects that lies beyond pair-wise summations.\cite{DobsonPerspective}

The vdW-DF method shares with RPA an electron-based foundation
as they avoid partitioning into, for example, atomic components. 
The methods also share a zero-point-energy counting nature (Sec.\ IV), 
an emphasis on approximating the ACF through longitudinal dielectrical 
functions that comply with the continuity equation (App.\ A), and conservation of 
the associated xc hole.\cite{dion,berland14p035412,bearcoleluscthhy14,ROPPreview}
One difference is that the vdW-DF  is based on a fully screened response 
description via a plasmon-based starting point that reflects a GGA-type 
internal-functional xc hole, whereas in RPA one starts with independent-particle 
excitations.\cite{Furche08,Eshuis12,Ren12p7447} With the full $E_c^{\nl}$ 
expression (\ref{eq:nonlocfunctionaldef}), used in an early seamless
vdW-DF functional,\cite{ryluladi00,rydberg03p126402} the vdW-DF method 
relies on the same machinery as RPA for systematically including 
screening effects, namely the Dyson equation for the density-density 
correlation function,\cite{lape77} as shown in Sec.\ II.

At the same time, the popular general-geometry vdW-DF versions and closely 
related variants\cite{dion,lee10p081101,cooper10p161104,vdwsolids,berland14p035412,hamada14}
rest on a second-order expansion (\ref{eq:Enlc_Slcnew}) of $E_c^{\nl}$, a step that 
is not used in RPA calculations.  An interesting question is then how 
much of the screening, collectivity, and nonadditivity effects are retained after 
making this truncation. The question is complex and we limit the discussion to 
making some comments in the context of a recent perspective article by 
Dobson.\cite{DobsonPerspective}

Dobson classifies nonadditivity effects as follows: Class `A' 
contains the effects of bond formation. These effects are automatically included 
in vdW-DF.  Class `B' is the spectator effect; that is, the modification by an
additional molecular-type fragment on the electrodynamics coupling between 
two molecular-type fragments.  Class `C' contains many-body effects that result 
with nondegenerate electron states and their ability to enhance the electronic response. 
A consequence is, for example, different asymptotic scaling laws for the vdW attraction
between sheets of metals or among metallic nanotubes than between
insulators.\cite{BarashI,BarashII,SerneliusI,SerneliusII,Dobson06,Lebegue10,DoGoDisp12,DobsonMB14,DobsonPerspective}
Class `B' and `C' are expected to be of greater importance for asymptotic 
interactions than at binding separations where there are 
contributions from many plasmons.\cite{Lebegue10,behy13,DobsonMB14}

For a vdW-DF version to fully address nonadditivity effects 
of class 'B', it requires that the evaluation proceeds with the full
interaction form,\cite{ryluladi00,rydberg03p126402} not the expansion 
(\ref{eq:Enlc_Slcnew}) used for the more recent and popular vdW-DF versions.\cite{dion,lee10p081101,berland14p035412}
However, vdW-DF reflects multipole enhancements in the binding among
molecules\cite{berland10p134705,berland11p1800} and image-plane formation 
in the binding of carbon nanotubes\cite{kleis08p205422} and in
challenging physisorption problems.\cite{lee11p193408,lee12p424213} 
Image-plane effects are captured in those recent expanded vdW-DF versions through 
the stronger sensitivity to the low-density regions arising at surfaces than 
to the high-density regions of the bulk.\cite{behy13} 

To fully capture effects in class 'C' one would also need to refine the 
vdW-DF inner-functional response model beyond a simple plasmon model 
relying on a GGA-based account.  However, some of the energetic impact of 
these effects is also, in practice and at a cruder level, reflected in 
the modern vdW-DF versions: In low-density, highly homogeneous systems, 
typical of a metal surface,\cite{lee11p193408,lee12p424213} the vdW-DF plasmon 
model yields small excitation energies, strongly enhancing non-local correlation 
effects.  On the other hand, except at edge regions, the GGA-based construction of 
vdW-DF does not distinguish between a limited molecular-type fragment, such as the center 
part of a polyaromatic hydrocarbon, and an extended fragment that has
no gap, such as graphene or a metallic nanotube.
This is a distinction that becomes important at asymptotic separations 
between fragments.\cite{Dobson06,Lebegue10,DoGoDisp12}

The recent vdW-DF versions\cite{dion,lee10p081101,berland14p035412} 
are formulated with the expectation that the second-order expansion (\ref{eq:Enlc_Slcnew}) 
is often sufficient in binding situations, with two molecular-type
fragments in close proximity.\cite{discussHollow} At binding it 
is important to treat truly nonlocal correlation effects and
the more local/semi-local correlation effects on a same
footing.\cite{rydbergthesis,dion,thonhauser,behy13} In the sections above 
we have illustrated that vdW-DF should not be viewed merely as a summation of 
contributions from pairs of density points. It is rather an expression of 
coupling of semilocal xc holes with a finite extension and with a shape and 
dynamics that already reflect a GGA-type response behavior.  The internal 
functional xc holes express a collectivity that generally extends beyond 
that of a single atom and represents a GGA-level of screening that we assume
is often adequate for treating interfragment binding.

The good performance of vdW-DF, in particular for the most recent nonempirical 
versions and variants, indicates that the vdW-DF method is capable 
of accurately reflecting the complicated balance that can exist between general
interaction contributions.\cite{lee11p193408,lee12p424213,behy13,bearcoleluscthhy14} 
For example, the recent consistent-exchange vdW-DF-cx version\cite{berland14p035412} 
can correctly describe the competition between covalent and ionic bonds in 
ferroelectrics, and between exchange effects and ionic and vdW attraction 
in weak-organic chemisorption.\cite{bearcoleluscthhy14} One needs 
Axilrod-Teller\cite{ra,Axilrod} corrections
(\ref{eq:partitionedACFstartrio}) and beyond\cite{ryluladi00,rydberg03p126402,Furche08,Tkatchenko12} 
to fully characterize the general dispersive interaction in systems that have three 
or more molecular-type regions. However, when two molecular-type regions are at their 
binding separation there is not generally room for a third molecule to get close and 
significantly influence that coupling. 

Finally, for a quantitative discussion of what screening effects 
are retained in recent vdW-DF versions one needs to compare 
the results of these versions with those obtained when using the 
full $E_c^{\nl}$ form\cite{ryluladi00,rydberg03p126402} under the
same approximation for the internal-functional description. Two of us have 
led one early such exploration\cite{SvetlaCompare} but no conclusion 
can be reached in that study because we refined the plasmon model between 
the layered-geometry formulation of Ref.\ [\onlinecite{rydberg03p126402}] 
and the launching of vdW-DF1.\cite{dion} A comparison of the results
based on the modern $S_{xc}$ response form is 
beyond the present scope.

\section{Summary}
\label{sec:summary}

Several formal properties of the vdW-DF theory have been highlighted. 
Specifically, we have documented how an effective internal xc hole can be 
viewed as a central building block in obtaining the nonlocal correlation of vdW-DF. 
We have documented how this internal xc hole resembles the xc hole construction 
that underpins standard GGA descriptions. 
Further, we have argued how the nonlocal correlations in vdW-DF can be interpreted as arising from the shift in collective modes induced by the electrodynamical coupling between such xc holes. 
This argument connects vdW-DF to the well-established RA picture of vdW-DF 
interactions.  Finally we have compared the vdW-DF method to RPA.

By discussing the formal properties of vdW-DF and links to other theories, we hope to help build bridges that stimulate the dissemination of ideas, not just within the field of van der Waals interactions but also within the wider field of material modeling.

\begin{acknowledgments}
The authors are grateful for many insightful discussions with
David C.\ Langreth, who passed away in 2011, and with
Bengt I.\ Lundqvist.  The authors are furthermore grateful for 
input, insight, and encouragement from Prof.\ G.~D.\ Mahan, primarily 
during his two extended visits to Chalmers, in 2009 and 2011.  This 
work was supported by the Swedish research council (VR) and the Chalmers 
Materials Area of Advance theory initiative.
\end{acknowledgments}
\begin{appendix}
\section{Role of the continuity equation in vdW-DF}
\label{sec:longproj}
This appendix details the correct longitudinal projection
in electrodynamics and is based on notes by and discussions with 
D.C. Langreth. It serves to further motivate writing the xc energy
\begin{equation}
E_{xc} +E_{\self} = \int_0^{\infty} \, \frac{du}{2\pi} \,
\hbox{Tr} \{ \ln(\nabla \epsilon(\im u) \cdot \nabla G) \}
\label{eq:excorenergyprojected}
\end{equation}
so that it expresses the exact longitudinal projection. 
The form (\ref{eq:excorenergyprojected}) reflects the 
continuity equation as well as the constituent equations of 
the electrodynamical response in materials.

Consider Ohm's law for the current $\mathbf{j}^{\rm ind}$ induced by 
a local field $\mathbf{E}=-\nabla \Phi_{\rm loc}$,
\begin{equation}
\mathbf{j}^{\rm ind}(\rr,\omega) = \int \, 
d\rrm \bm{\sigma}(\rr,\rrm,\omega)  \mathbf{E}(\rrm,\omega) \, ,
\label{eq:jinduced}
\end{equation}
and corresponding to the induced charges $\rho^{\rm ind}$. In \eq{eq:jinduced},
we use $\bm{\sigma}(\rr,\rrm,\omega)$ to denote the nonlocal conductivity 
tensor.  In turn, this tensor corresponds to a nonlocal dielectric function 
\begin{equation}
\bm{\varepsilon}(\rr,\rrm,\omega) = \mathbf{1} + \frac{4\pi \im}{\omega} 
\bm{\sigma}(\rr,\rrm,\omega) \, .
\end{equation}

Fourier transforming gives $\mathbf{E}_{\mathbf{k}}= -i \mathbf{q}\, \Phi_{{\rm loc},\mathbf{q}}$,
and a continuity specification
\begin{equation}
\omega \rho_{\mathbf{q}}(\omega) = \mathbf{q} \cdot j^{\rm ind}_{\mathbf{q}} \, ,
\end{equation}
that relates the local field and the induced charge
\begin{eqnarray}
4\pi \rho^{\rm ind}_{\mathbf{q}}(\omega) 
& = & - \mathbf{q} \cdot \sum_{\mathbf{q}'} \left[ 
\bm{\varepsilon}_{\mathbf{q},\mathbf{q}'} (\omega) - \mathbf{1}\delta_{\mathbf{q},\mathbf{q}'}\right] \cdot
\mathbf{q}\, \Phi_{{\rm loc},\mathbf{q}} \, .
\end{eqnarray}
We infer an exact, general microscopic relation between the (longitudinal) local-field
response $\tilde{\chi}$ and the dielectric tensor 
\begin{eqnarray}
4\pi \tilde{\chi}_{\mathbf{q},\mathbf{q}'} & = & - \mathbf{q} \cdot 
\left[ \bm{\varepsilon}_{\mathbf{q},\mathbf{q}'} - \mathbf{1} \delta_{\mathbf{q},\mathbf{q}'} \right]
\cdot \bm{q}' \, ,
\label{eq:microscopicrelationFourier}
\\
\tilde{\chi} & = & \frac{1}{4\pi} \nabla \cdot (\bm{\varepsilon} - \mathbf{1}) \cdot \nabla  \, .
\label{eq:microscopicrelation}
\end{eqnarray}
The result \eq{eq:microscopicrelationFourier} reflects the continuity equation and 
identifies
\begin{equation}
{\bm{\varepsilon}}^{\rm long}_{\mathbf{q},\mathbf{q}'} \equiv \hat{\mathbf{q}}
\cdot {\bm{\varepsilon}}_{\mathbf{q},\mathbf{q}'}\cdot 
\hat{\mathbf{q}}' \, ,
\label{eq:microscopicrelationFourierFin}
\end{equation}
as the proper (consistent) definition of the longitudinal projection 
of the dielectric repose in an inhomogeneous system. 

In \eq{eq:microscopicrelation} the difference $(\bm{\varepsilon} - \mathbf{1})/4\pi$ 
takes the form of a local-field susceptibility tensor $\bm{\sigma}$. Using
$\bm{\sigma}_{\rm ext}$ to denote the corresponding external-field susceptibility
we also have a relation for the external-field response
\begin{equation}
\chi = \nabla \cdot \bm{\sigma}_{\rm ext} \cdot \nabla  \, .
\label{eq:microscopicrelationForExtField}
\end{equation}

Finally, the correct longitudinal projection of the dielectric tensor is given by
\begin{equation}
\kappa \equiv \nabla \cdot \bm{\varepsilon} \cdot \nabla G = \varepsilon^{\rm long}
\, .
\label{eq:longprojectdieletrics}
\end{equation}
This is demonstrated by expressing the microscopic relation in wavevector space
\begin{eqnarray}
4\pi\tilde{\chi}_{\mathbf{q},\mathbf{q}'} & = & - \mathbf{q}\cdot \left[ {\bm{\varepsilon}}_{\mathbf{q},\mathbf{q}'} -\mathbf{1} \delta_{\mathbf{q},\mathbf{q}'} \right]\cdot \mathbf{q}'
\nonumber \\
& = & - |\mathbf{q}||\mathbf{q}'| {\bm{\varepsilon}}^{\rm long}_{\mathbf{q},\mathbf{q}'} + q^2 \delta_{\mathbf{q},\mathbf{q}'} \, , 
\end{eqnarray}
where we have used the $\varepsilon^{\rm long}$ 
specification \eq{eq:microscopicrelationFourierFin}.
Solving for ${\bm{\varepsilon}}^{\rm long}_{\mathbf{q},\mathbf{q}'}$ gives 
\begin{equation}
{\bm{\varepsilon}}^{\rm long}_{\mathbf{q},\mathbf{q}'} = \delta_{\mathbf{q},\mathbf{q}'} - \frac{4\pi}{|\mathbf{q}||\mathbf{q}'|} 
\tilde{\chi}_{\mathbf{q},\mathbf{q}'} \, .
\end{equation}

\end{appendix}


\begin{thebibliography}{99}
\bibitem{notationfootnote} All response functions and dielectric
functions are nonlocal, i.e., functions of two spatial coordinates.  
The dependence is generally suppressed by considering equivalent operators
$A(\omega;\mathbf{r},\mathbf{r}') \equiv \langle \mathbf{r} | A(\omega) | \mathbf{r}'\rangle$.
The Fourier transform is defined as $A(\omega;\mathbf{q},\mathbf{q}') 
\equiv \int \, d\mathbf{r} \int \, d\mathbf{r}' \, 
        e^{-i\bq\cdot\mathbf{r}} \,
        \langle \mathbf{r} | A(\omega) | \mathbf{r}'\rangle\,
        e^{i\bqm\cdot\mathbf{r}'}.$
\end{thebibliography}
\end{document}